\documentclass[conference]{IEEEtran}
\IEEEoverridecommandlockouts
\usepackage{cite}
\usepackage{amsmath,amssymb,amsfonts}
\usepackage{graphicx}
\usepackage{textcomp}
\usepackage{xcolor}
\usepackage[square,numbers]{natbib}

\newcommand{\regtm}{\texttrademark}
\usepackage{url}
\usepackage{mathrsfs}

\usepackage[endLComment=]{algpseudocodex}
\usepackage[ruled]{algorithm}

\usepackage{booktabs}
\usepackage{adjustbox}
\newcommand{\AlgIn}{\Statex \textbf{Input:} }
\newcommand{\AlgOut}{\Statex \textbf{Output:} }

\usepackage{latexsym}
\usepackage{verbatim}
\usepackage{balance}

\usepackage{multicol}

\usepackage[c3, nocomma, short]{optidef}

\usepackage{xspace}
\usepackage{enumitem}

\makeatletter 
\newcommand{\linebreakand}{%
  \end{@IEEEauthorhalign}
  \hfill\mbox{}\par
  \mbox{}\hfill\begin{@IEEEauthorhalign}
}
\makeatother 

\def\BibTeX{{\rm B\kern-.05em{\sc i\kern-.025em b}\kern-.08em
    T\kern-.1667em\lower.7ex\hbox{E}\kern-.125emX}}
\begin{document}

\title{Anonymized Network Sensing \\ using \texttt{C++26 std::execution} on GPUs }

\author{\IEEEauthorblockN{Michael Mandulak}
\IEEEauthorblockA{\textit{Rensselaer Polytechnic Institute} \\
Troy, NY, USA\\
mandum@rpi.edu}
\and
\IEEEauthorblockN{Sayan Ghosh}
\IEEEauthorblockA{\textit{Pacific Northwest National Laboratory} \\
Richland, WA, USA \\
sayan.ghosh@pnnl.gov}
\and
\IEEEauthorblockN{S M Ferdous}
\IEEEauthorblockA{\textit{Pacific Northwest National Laboratory} \\
Richland, WA, USA\\
sm.ferdous@pnnl.gov}

\linebreakand

\IEEEauthorblockN{Mahantesh Halappanavar}
\IEEEauthorblockA{\textit{Pacific Northwest National Laboratory} \\
Richland, WA, USA\\
hala@pnnl.gov}
\and
\IEEEauthorblockN{George Slota}
\IEEEauthorblockA{\textit{Rensselaer Polytechnic Institute} \\
Troy, NY, USA\\
slotag@rpi.edu}
}

\maketitle

\thispagestyle{plain}
\pagestyle{plain}

\begin{abstract}
Large-scale network sensing plays a vital role in network traffic analysis and characterization. As network packet data grows increasingly large, parallel methods have become mainstream for network analytics. While effective, GPU-based implementations still face start-up challenges in host-device memory management and porting complex workloads on devices, among others. To mitigate these challenges, composable frameworks have emerged using modern C++ programming language, for efficiently deploying analytics tasks on GPUs. Specifically, the recent C++26 \texttt{Senders} model of asynchronous data operation chaining provides a simple interface for \emph{bulk pushing tasks} to varied device execution contexts.  

Considering the prominence of contemporary dense-GPU platforms and vendor-leveraged software libraries, such a programming model consider GPUs as first-class execution resources (compared to traditional host-centric programming models), allowing convenient development of multi-GPU application workloads via expressive and standardized asynchronous semantics. In this paper, we discuss practical aspects of developing the Anonymized Network Sensing Graph Challenge on dense-GPU systems using the recently proposed C++26 \texttt{Senders} model. Adopting a generic and productive programming model does not necessarily impact the critical-path performance (as compared to low-level proprietary vendor-based programming models): our commodity library-based implementation achieves up to 55$\times$ performance improvements on 8$\times$ NVIDIA A100 GPUs as compared to the reference serial GraphBLAS baseline. 

\end{abstract}


\begin{IEEEkeywords}
Network traffic analysis, Multi-GPU, C++26, \texttt{std::execution}, Graph Challenge
\end{IEEEkeywords}

\section{Introduction}\label{sec:intro}
Network sensing at a large-scale \cite{Kepner_2023} plays a critical role in a variety of applications, such as in network traffic characterization, recommender systems, sensor dynamics and network data analytics \cite{NetSenseSurvey,Kawaminami_2022,RecommenderSys}. This is further emphasized through the recently proposed Anonymized Network Sensing Graph Challenge \cite{graphchallenge}, highlighting contributions to the generation and analysis pipeline of anonymized traffic matrix data from network packet captures (in the form of a standardized file format, namely, PCAP). Focusing on the analysis portion of this Graph Challenge, GPU-based analytics have become increasingly relevant within large-scale network data workloads \cite{gpuNetwork1,gpuNetwork2,gpuNetwork3}. Consequently, dense-GPU systems (i.e., 4--16 GPUs on a single node) are commonplace, with terabytes of aggregate memory and proprietary GPU interconnection network to enhance the overall bandwidth. However, designing codes to exploit several GPUs must contend with the challenges associated with effective workload distribution, host-device memory traffic, GPU interconnection performance, and complex algorithms \cite{bigDataNetSurvey}, to name a few. 
\begin{figure}[!ht]
    \centering
    \includegraphics[scale=0.29]{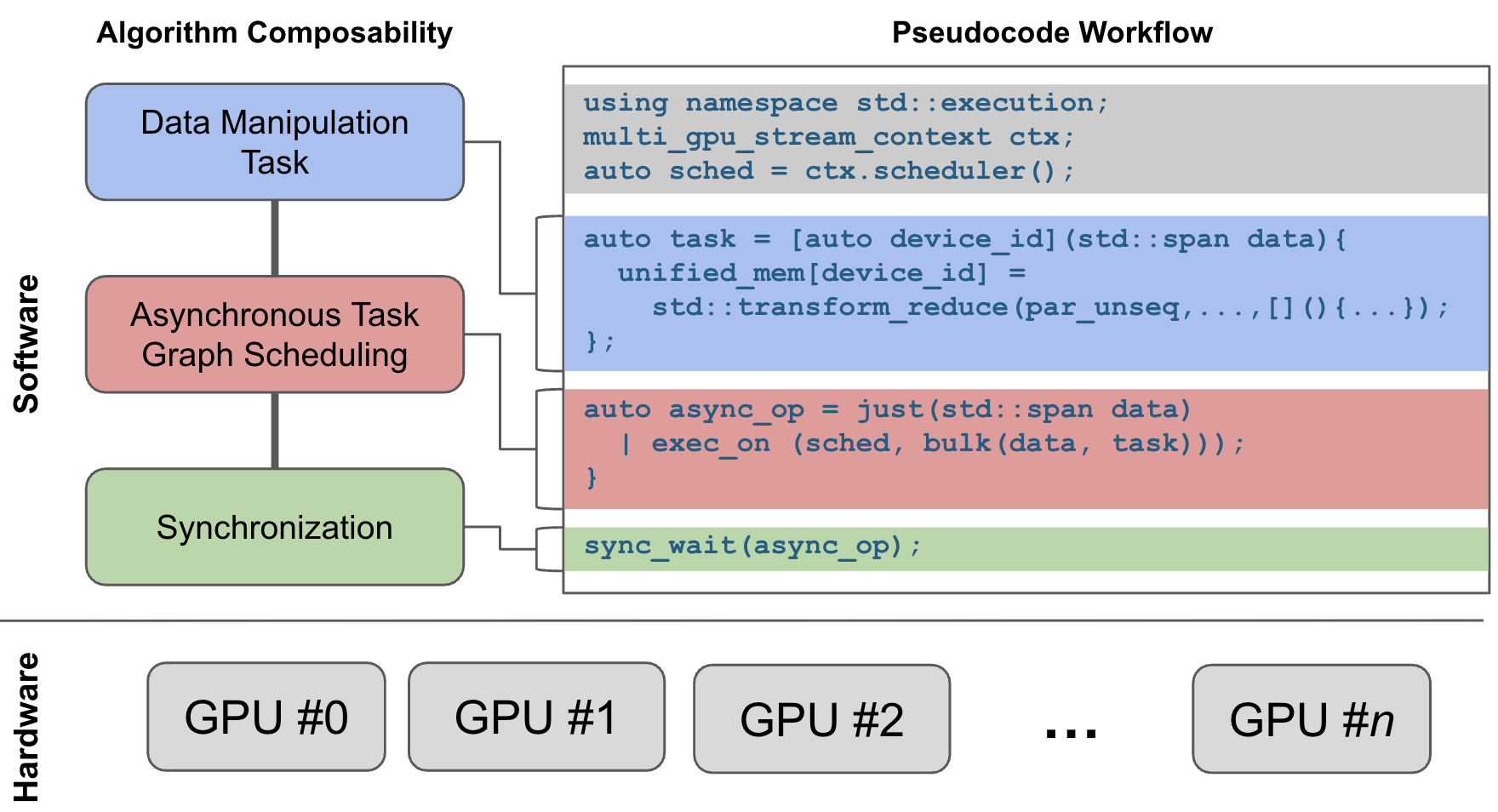}
    \caption{C++26 asynchronous \texttt{Senders} programming model in \texttt{std::execution}. Using this standardized model, asynchronous workloads can be composed over diverse execution environments, for e.g., scheduling data manipulation tasks on multiple GPUs, captured here.}
    \label{fig:prog_model}
\end{figure}
These challenges were traditionally mitigated through vendor-based library solutions coupled with implementations explicitly composing the underlying data structures and associated operations. However, to enhance the portability of workloads targeting contemporary device\slash accelerator environments, driving the development using vendor-agnostic standardized programming models is key. Towards that goal, the C++ programming language has made significant inroads in improving performance and productivity to span across a wide variety of applications and hardware spectrum. C++26 is the ``next'' generation of the C++ standard (considered a landmark release since C++11 ushering in the modern C++ era, a decade ago), including significant changes to the library itself relative to the core language features\footnote{\url{https://en.cppreference.com/w/cpp/compiler_support/26}}. Specifically, C++26 Execution control library (referred as \texttt{std::execution} or \texttt{Senders} framework) provides composable building blocks for managing asynchronous execution on generic execution environments, as depicted in Fig.~\ref{fig:prog_model}. This library allows transparent composition of task execution graphs or Directed Acyclic Graphs (i.e., DAGs similar to CUDA graphs\footnote{\url{https://developer.nvidia.com/blog/cuda-graphs/}}) and asynchronous scheduling on queues associated with specific execution contexts. It also introduces explicit syntax for launching the tasks (and invoking subsequent callbacks or ``receivers''), thus, providing enhanced flexibility for common ``many-tasks'' scenarios compared to the existing C++  \texttt{std::future} or \texttt{std::packaged\_task} interfaces.

These developments propose explicit high-level asynchronous semantics for GPU-based paradigms, allowing for the translation of ``many-tasks'' scenarios to many-threaded device contexts through vendor-optimized standardized library functions such as transform, reduce, scan, etc. Following this, we explore the efficacy of the C++26 \texttt{Senders} model on GPUs for the large-scale data analytics workload depicted in the Anonymized Network Sensing Graph Challenge. 

This paper is one of the early works to explore C++26 \texttt{std::execution} workflows within large matrix workloads on multiple GPUs. Our contributions are as follows:
\begin{itemize}
  \item Develop Anonymized Network Sensing Graph Challenge using recently proposed C++26 \texttt{Senders} model, leveraging data structures built on top of Standard Library containers for platform-agnostic analytics. 
    \item Incorporated generic data batching scheme to subpartition network data between host and GPU (for handling partition sizes exceeding the available GPU memory).
    \item Achieved performance improvements of up to 55$\times$ compared to the serial GraphBLAS-based reference implementation using 8$\times$ NVIDIA A100 GPUs.
\end{itemize}

\section{Background}\label{sec:background}

\subsection{Graph Challenge}
The Anonymized Network Sensing Graph Challenge \cite{graphchallenge} highlights contributions towards improved network traffic analytics through multiple facets. These include the improvement of PCAP I/O, packet data extraction methods, IP anonymization, traffic matrix construction, matrix storage, and matrix analytics. This work focuses on improving the final step of network traffic analytics through the application of dense-GPU processing. 
\begin{figure}[!ht]
    \centering
    \includegraphics[scale=0.32]{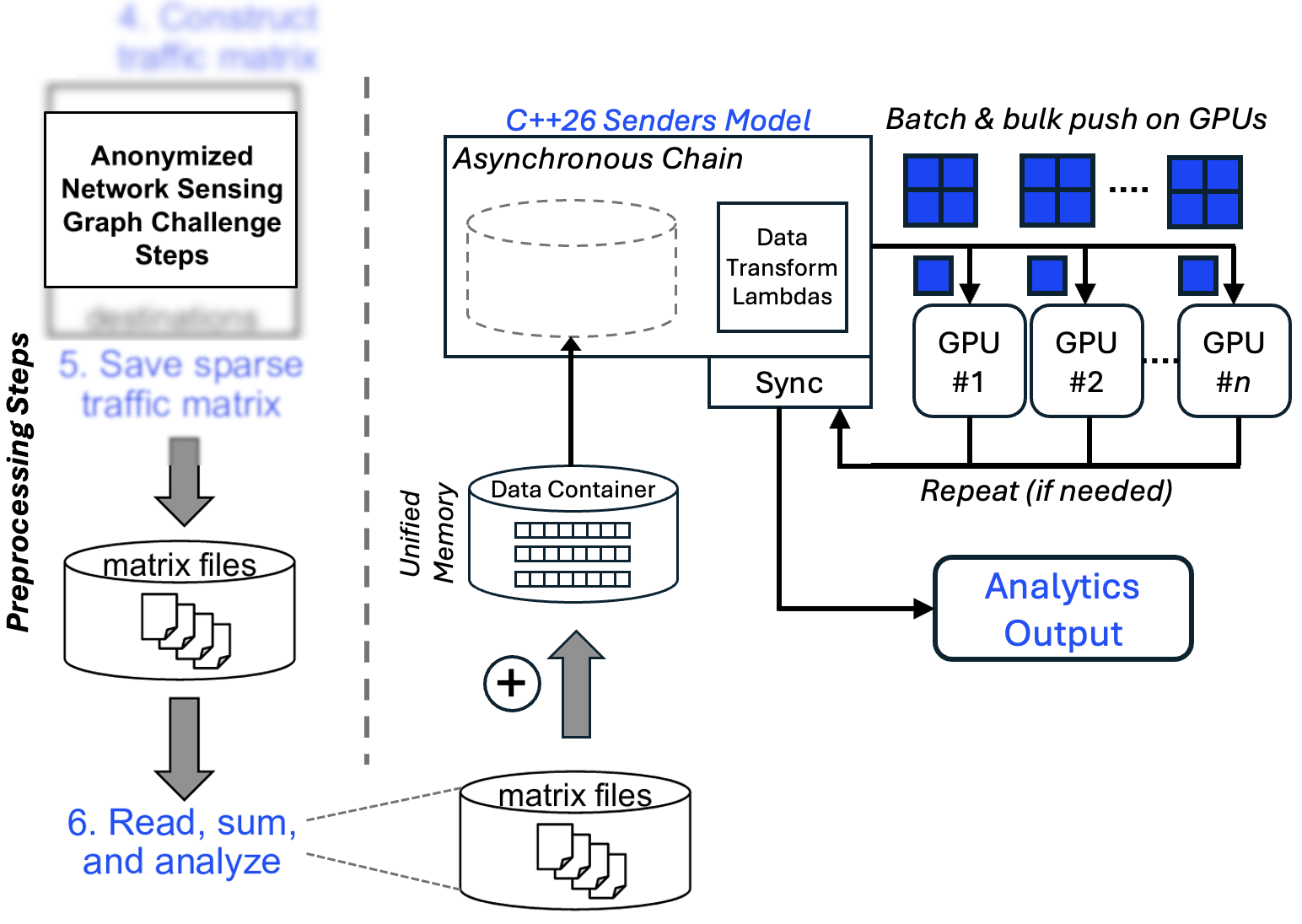}
    \caption{Graph Challenge workflow overview with our proposed processing method using the \textit{C++26 Senders Model}. The "Preprocessing Steps" graphics are adapted from \cite{graphchallenge}. Under the C++26 Senders Model, we load and aggregate traffic matrix files, forming an asynchronous workflow, comprised of batching and bulk pushing data operations to multiple GPUs.}
    \label{fig:GCoverview}
\end{figure}
At a high level, we outline our method in Fig.~\ref{fig:GCoverview}. We begin with the pre-processing steps adapted from the Graph Challenge proposal, extending the analysis portion to the loading of matrix files into generalized data containers. Using these, we follow propositions in the \textit{C++26 Senders Model} workflow, performing data analytics by pushing asynchronous chains of workloads to a multi-GPU setup and retrieving the analytics output. Next, we describe the components of the \textit{C++26 Senders Model}. 

\subsection{C++26 Execution Model}

Within the P2300 proposal for standard C++ \cite{p2300r10}, the authors propose an asynchronous task management workflow under the \texttt{std::execution} namespace for composable task-based parallel workloads launched on generic execution resources (i.e., an abstraction for a hardware component), handled by a scheduler. Overall, this proposal builds upon the popular asynchronous callback-promise workflow under a ``senders-receivers'' (\texttt{Senders}) model, associating an asynchronous unit of computation to a \emph{sender} with a completed state being processed under a \emph{receiver}.  The \emph{execution resource} of this model is expressed via an explicit scheduler abstraction; scheduler represents the place where execution happens (with options to compose senders into task graphs), which could be a thread-pool, single thread or GPU(s).

Vendor-specific implementations of \texttt{std::execution} are in active development to extend the \texttt{Senders} model to accelerators, such as GPUs. Specifically, NVIDIA is enhancing integration of proprietary CUDA with the C++ Standard Library (STL) functionalities on device through \texttt{libcudacxx} \cite{libcuxx}, backporting several libraries relative to data structures and operations under the \texttt{cuda::std} namespace. The integration of STL data structures and basic operations (such as \texttt{span, mdpsan, reduce,} etc.) is extended within the \texttt{Senders} model through \texttt{nvexec} \cite{nvexec}, NVIDIA's device-based schedulers and functionalities for the relevant asynchronous abstractions. There are some generalizations, such as a default allocation of all used device memory to unified memory under \texttt{nvexec}. Overall, the \texttt{stdexec} library is a reference implementation of the C++26 Senders model, whereas NVIDIA C++17 Standard Parallelism (\texttt{stdpar}) allows GPU acceleration of the standard algorithms, both are integrated within the NVIDIA HPC SDK suite.

\section{Methodology}\label{sec:method}
We introduce the key components in implementing network traffic analytics challenge using C++26 \textit{senders}-based programming model for containerized data processing, in \S\ref{ssec:method-programming} and \S\ref{ssec:method-containers}. In \S\ref{ssec:method-batching}, we discuss subpartitioning the input data for concurrent batch processing on GPUs. 

\subsection{Programming and Execution Model}\label{ssec:method-programming}
From the perspective of our workload, high-level components of the C++26 asynchronous senders model are captured in Fig.~\ref{fig:prog_model}. We first  distinguish between the hardware and software layers of this model, followed by task composition.

\textbf{Software Model:}
We first provide an overview of the pseudocode workflow for a generic transformation and reduction operation from Fig.~\ref{fig:prog_model}, making generalizations for conciseness. The model relies on the usage of a vendor-specific device scheduler built on top of a single or multi-device context (depicted as \texttt{sched} in Fig.~\ref{fig:prog_model}). Using this abstraction, programming on single vs. several devices (i.e., execution resources) are almost indistinguishable; we rely on a multi-GPU context to bulk push operations on a single dense-GPU node.

We consider the data manipulation tasks launched on devices as \emph{lambda} functions (lambda expressions or lambdas are unnamed functions used to define operations where they are invoked), with \texttt{std::span} to represent a non-owning view of the contiguous data for processing. This allows for application extensibility, as we can process data of arbitrary contiguous sequences (i.e., trivially copyable)through \texttt{span}. Since vendor extensions such as \texttt{nvexec} (which provides NVIDIA specific schedulers and runtime support) assumes unified memory (i.e., single shared address space between host and device), we perform operations (in this case, \texttt{transform\_reduce}) within the \texttt{std} namespace and commit results to unified memory relative to the device index (see Fig.~\ref{fig:prog_model}). 

The asynchronous data manipulation tasks are chained and associated with specific data spans. While supporting complex task workflows under \texttt{std::execution} for asynchronous chaining, for our purposes we can simply push bulk executions to multiple devices (reusing \texttt{sched}) and the set data manipulation lambda accordingly. We then perform a synchronization on this asynchronous chain, retrieving the result at the end of the workflow. This can be further customized with collective operations or other synchronization measures (e.g., multi-node scenarios).

\textbf{Execution Environment:}
Traditional device-based applications rely on vendor-specific programming models, with separate multi-node or multi-GPU abstractions. In this model, various options exist to instantiate the underlying execution resource (i.e., \texttt{sched} in Fig.~\ref{fig:prog_model}), comprising of multiple CPU\slash host threads, single GPU or dense-GPUs (all GPUs in a node), while keeping the asynchronous task descriptions uniform. With appropriate backend support, this model can be expanded to distributed implementations (over network) considering partitioned data. 

\textbf{Tasks Composability:}
The basis of the programming model relies on the generalization of any simple or complex parallel computation into three basic components: a data manipulation task, asynchronous data operations and bulk synchronizations. In Fig.~\ref{fig:prog_model}, we provide an example of a generic transform and reduction operation (which transforms each element in the container in place and then accumulates the results), a common data manipulation task in a myriad of data analysis workflows. For our purposes, we demonstrate extensibility to common matrix-based network analysis measures, such as link counts, source counts and max degrees, among others. For more complex algorithms (i.e., which cannot be recast into a combination of containerized scan, reduce and transform operations), GPU integration with \texttt{stdpar} allows inclusion of device kernel within the data manipulation lambdas, allowing greater flexibility. 

\subsection{Standardized Containers}\label{ssec:method-containers}
The usage of \texttt{std::span} within the programming model allows expansive data backends for portable analytics. Popular frameworks, such as cuGraph \cite{cugraph} or Gunrock \cite{wang2016gunrock}, considers variety of input data formats (dedicated data ingestion frameworks such as cuDF~\cite{cudf} can alleviate preprocessing overheads), but often resort to containerized operations internally for implementing workloads. This method espouses the composable and generic aspects of C++ metaprogramming. Using this programming model, we adhere to similar principles, using \texttt{std::vector} container for the relevant data structures and passing an \texttt{std::span} for device invocations. Complex sparse representations, such as Compressed Sparse Row (CSR), can be managed through a collection of \texttt{std::vector} containers (e.g. consider a graph as a vector of a vector of vertex neighborhoods; operations such as triangle counting become a sequence of set intersections).

\subsection{Concurrent Batching}\label{ssec:method-batching}
Regarding data distribution across devices using multi-device schedulers, the usage of unified memory simplifies data access and allocation using on-demand page migration~\cite{sakharnykh2018everything}. This, however, assumes a simple device-data partitioning occurs to split data evenly among devices. For our purposes, we rely on this even split of matrix data per device, which allows us to further take advantage of data \textit{batching} on device. 

\begin{figure}[!ht]
    \centering
    \includegraphics[scale=0.29]{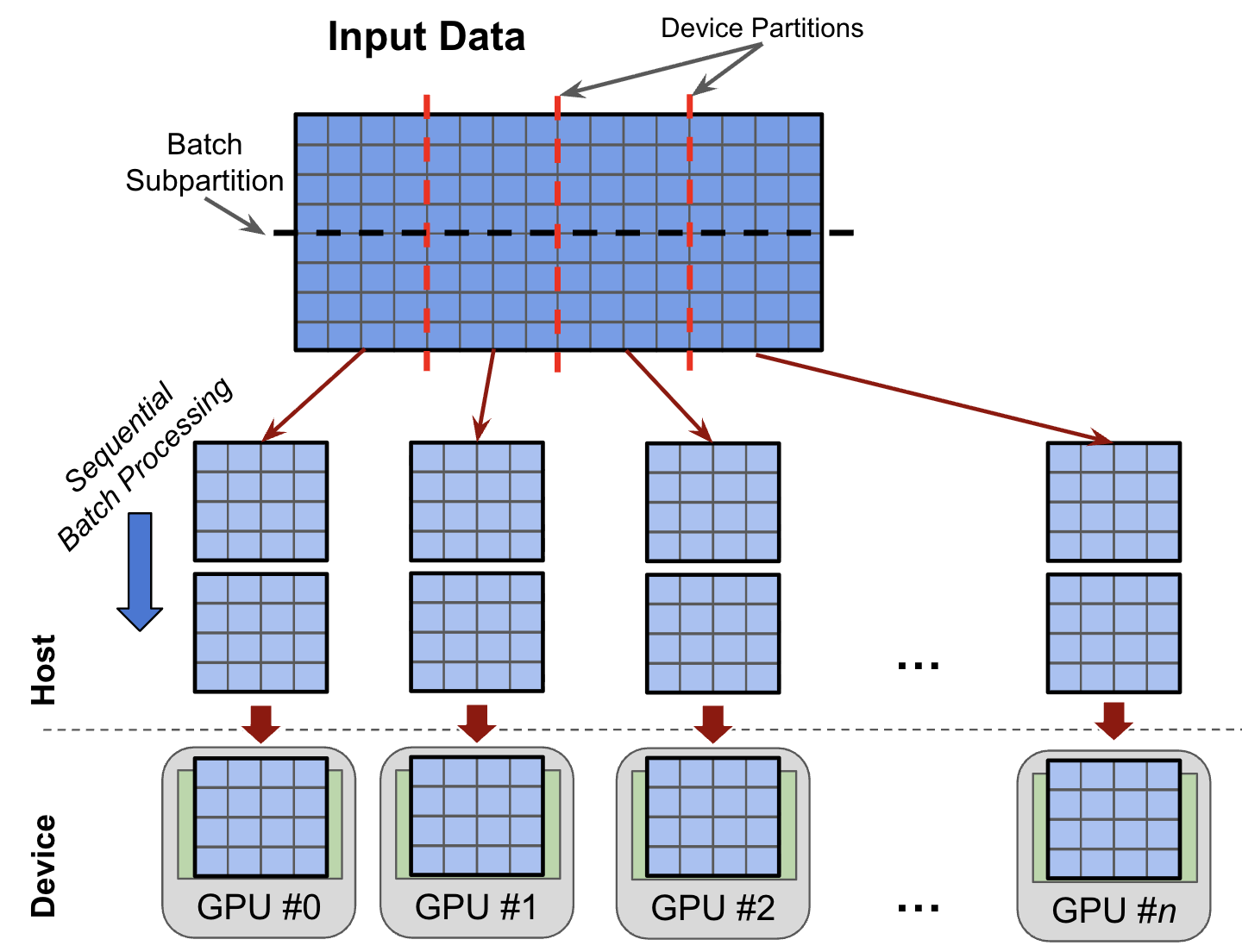}
    \caption{Demonstrates input data batching from host to GPUs, where individual device partitions are sub- partitioned into fixed-size batches, sequentially moved on GPUs by the host during computation.}
    \label{fig:batching}
\end{figure}

We consider \textit{batching} as a technique to subpartition input data into $b_n$ batches, where each batch is processed in parallel one after the other.  We detail this workflow in Fig.~\ref{fig:batching}. The batch count $b_n$ is input specified, where we assume $b_n=1$ is the default case (where the batch is the entire device data partition), and $b_n = k$ splits the input data into $k$ portions per device. This allows us to leverage sequential processing for better scalability and lower page faults through reduced data sizes at processing time. We further discuss the impact on performance in \S\ref{sec:results}.

\section{Implementation Details}\label{sec:implementation}
In this section, we discuss our implementation alongside the design choices for Graph Challenge analytics, listed in Table~\ref{tab:properties}.

\begin{table}[h]
\centering

\caption{Graph Challenge packet analysis measures, with the aggregate properties and summation notations adapted from \cite{graphchallenge}. The relevant programming model data operation is listed for each property. $A_t$ represents a network traffic matrix at time $t$, with $A_t(i,j)$ as the number of packets between source $i$ and destination $j$. }
\label{tab:properties}
\begin{tabular}{@{}lll@{}}
\toprule
\textbf{Aggregate Property}     & \textbf{Notation} & \textbf{C++ function} \\ \midrule
Valid packets $N_V$                  & $\sum_i \sum_j A_t(i, j)$ & \texttt{reduce}($weights$) \\
Unique links                         & $\sum_i \sum_j \left| A_t(i, j) \right|_0$ & \texttt{size}($edges$) \\
Unique sources                       & $\sum_i \left| \sum_j A_t(i, j) \right|_0$ & \texttt{size}($row\_sums$) \\
Max source fan-out      & $\max_i \left| \sum_j A_t(i, j) \right|_0$ & \texttt{max}($degrees$) \\
Unique destinations                  & $\sum_j \left| \sum_i A_t(i, j) \right|_0$ & \texttt{size}($col\_sums$) \\
Max destination fan-in  & $\max_j \left| \sum_i A_t(i, j) \right|_0$ & \texttt{max}($degrees$) \\
\bottomrule
\end{tabular}
\end{table}

\subsection{Software Dependencies}
We utilize a number of vendor-based libraries that integrate C++26 \texttt{Senders} model execution workflows on GPUs, specifically NVIDIA CUDA-based implementations of \texttt{std::execution} workflows through \texttt{nvexec} \cite{nvexec} for the translation of asynchronous workflows and operation lambdas to device code. \texttt{nvexec} also provides the multi-GPU scheduler context used to designate the execution resource. Our containerized data structures are STL-based and uses \texttt{libcudacxx} or \texttt{libcu++} \cite{libcuxx} for automatic unified memory management. We further use \texttt{libcu++} for device implementations of the STL operations, such as \texttt{std::reduce} available in \texttt{cuda::std} namespace.

\begin{algorithm}[t]
\caption{Pseudocode for \emph{max reduction}}\label{alg:max_red}
\begin{algorithmic}[1]
\AlgIn{Container: $data$, Multi-GPU Context: $ctx$,} \hspace{0.5cm}Batch Count: $b_n: n \in \mathbb{R}_{>0}$
\AlgOut{Maximum value}
\Procedure{Max\_Lambda}{\texttt{cuda::std::span}($data$), \texttt{cuda::std::span}($result$}
    \State $d \gets$ device\_id
    \State $result[d] \gets$ \texttt{max}($result[d]$, \label{algLine:lambda_op} 
    \State \; \; \texttt{cuda::std::reduce}($data,$ \texttt{std::max}))  \label{algLine:stdfunc}
\EndProcedure

\State $sched \gets$ $ctx$\sc{.get\_scheduler()} 
\State \texttt{cuda::std::span} $result \gets \emptyset$
\For{each batch $b$ of $data$} 
    \State $subspan \gets $ \texttt{cuda::std::span}($b$)
    \State $sndr \gets $ \texttt{stdexec::just}($subspan, result$)
    \State $|$ \texttt{exec\_on}($sched$, \texttt{stdexec::bulk}($size(b),$
    \State \hspace{2cm}\sc{Max\_Lambda}) 
    \State \texttt{stdexec::sync\_wait(std::move}($sndr$))
\EndFor
\State \Return $result$

\end{algorithmic}
\end{algorithm}
\subsection{Data Representation}

In the processing of the network traffic matrix files, we refer to the sequential GraphBLAS-based implementation discussed in \cite{graphchallenge}. Instead of GraphBLAS data representations, we use generic STL-based containers, requiring intermediate transformation of the CSR into flat containers comprising of the $edges$, $degrees$ and $weight$ data for the respective nonzero entries. Using these source containers, we can build derivative containers for specific data operations, such as $row\_sums$ and $col\_sums$ for out-- and in--degrees, respectively.

\subsection{Analytics and Operations}

The associated analytics include source, destination and total packet counts, maximum fan-in/fan-out and maximum link counts. For all the required measures, a combination of sum reductions and maximum scans suffice, of which we generalize for repeated data processing among the containers. The Graph Challenge properties are listed in Table \ref{tab:properties}, alongside data manipulation operations used to implement the specific functionality. Each operation can be successively invoked on subsequent batches of data, analyzing sub-partitioned portions of the packet data on device (this option is discussed in \S\ref{ssec:method-batching}). Details of the two primary operations (i.e., scan and reduce) are as follows: 

\textbf{Maximum Scan:} A concise example of the maximum reduction implementation is listed in Pseudocode~\ref{alg:max_red}, represented as the lambda expression \textsc{Max\_Lambda}, which performs maximum reduction on device using \texttt{cuda::std::reduce} on a unified memory-backed \texttt{span}. The data can be optionally processed in more than one batches, designating an even number of sub-partitions of the data relative to the number of batches in $subspan$. The asynchronous tasks are encapsulated through the \emph{sender} $sndr$, passing the input span and result. Next, we specify the resource context (passing the scheduler and lambda) on which the \emph{sender} will be executed (multiple senders can be successive chained\slash scheduled, i.e., utilizing overloaded \texttt{operator|}). Finally, a blocking synchronization on sender completes the operation.

\textbf{Sum Reduction:} For similar reduction or single buffer type operations, we follow the exact same structure as in Pseudocode~\ref{alg:max_red}. We can simply replace the data manipulation operation in Lines \ref{algLine:lambda_op} and \ref{algLine:stdfunc} with the relevant operation, and perform the same workflow.

\section{Evaluations}\label{sec:results}
 We first describe our software and hardware setup, followed by execution time performance results and an overall analysis. We compare our implementation to the sequential GraphBLAS-based Python implementation provided as part of the Graph Challenge \cite{graphchallenge}.

\textbf{Platform:}
Our primary GPU platform in our evaluations is an NVIDIA\regtm{} DGX system with 8 GPUs (DGX-A100). The DGX A100 is the third generation server node from NVIDIA\regtm{}, and consists of 8 ``Ampere'' A100 GPUs (with 108 SMs) with 40GB HBM2 memory\slash GPU and two-way 64-core AMD EPYC 7742 CPUs at 2.25GHz, 256MB L3 cache, 8 memory channels, and 1TB DDR4 memory. NVIDIA GPUs either come in the PCIe or proprietary NVLink\slash NVSwitch based form factors. Proprietary SXM module allows NVIDIA GPUs to directly communicate through NVLink interconnect (DGX-A100 uses SXM4).

Our implementation is built using CUDA version 12.2, GCC version 13.3.0 and version 25.5-0 of the NVIDIA High Performance Computing Software Development Toolkit\footnote{\url{https://developer.nvidia.com/hpc-sdk}} for \texttt{nvc++}. We use \texttt{std=c++20}, as C++26 features are experimental using the \texttt{--experimental-stdpar} flag. Experimental NVIDIA device support features are available using the \texttt{-stdpar=gpu} flag. Our source code is openly available from GitHub: \url{https://github.com/mmandulak1/stdexecANSGC}.

\textbf{Dataset:}
We use the specific data required for this Graph Challenge \cite{graphchallenge}, which are GraphBLAS matrices derived from randomized network packet data of $2^{30}$ synthetic packets.


\subsection{Baseline Execution}\label{ssec:baseline}
\begin{figure}[!ht]
    \centering
    \includegraphics[scale=0.54]{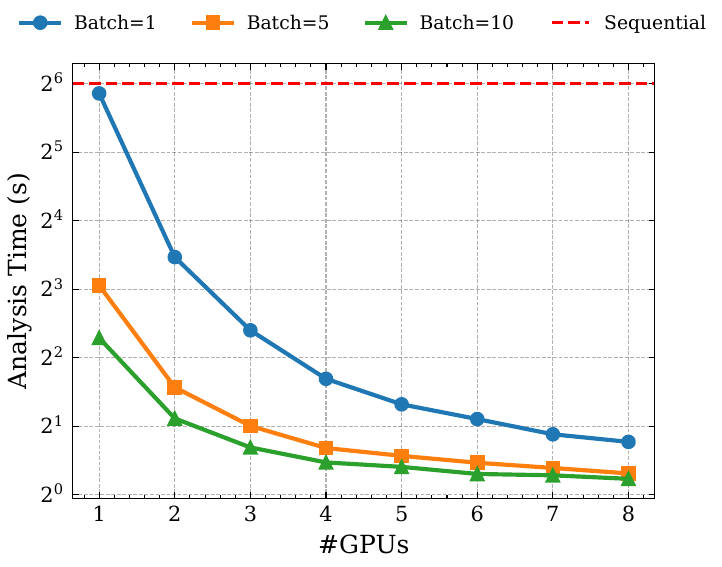}
    \caption{Scalability (analysis time, \emph{lower} is \emph{better}) on 1--8 GPUs with varying batch counts. Best performance is observed using 8 GPUs and 10 batches at $\sim$1 seconds compared to $\sim$64 seconds for sequential baseline.}
    \label{fig:batch_analysis}
\end{figure}

\begin{figure}
    \centering
    \includegraphics[scale=0.54]{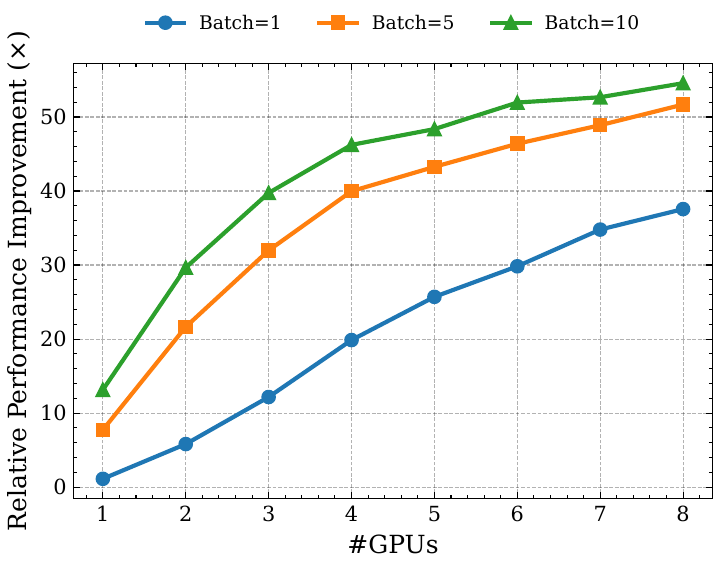}
    \caption{Relative performance improvement (compared to serial reference implementation, \emph{higher} is \emph{better}) on 1--8 GPUs with varying batch counts. Best performance observed at 8 GPUs using 10 batches: 55$\times$.}
    \label{fig:batch_analysis_speedup}
\end{figure}

\begin{figure}
    \centering
    \includegraphics[scale=0.54]{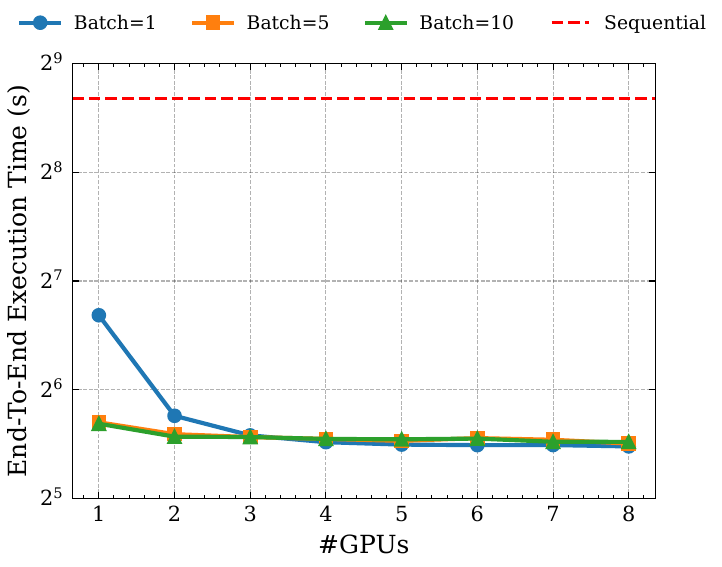}
    \caption{Scalability (end-to-end time, \emph{lower} is \emph{better}) on 1--8 GPUs with varying batch counts. Best performance observed using 8 GPUs and default batch count of 1, leading to 9$\times$ improvement vs. sequential.}
    \label{fig:batch_total}
\end{figure}

We separate baseline execution time assessments into two tasks: \emph{analysis time} and \emph{end-to-end time}. Analysis time is the time taken to calculate the analysis measures, excluding preprocessing (e.g., data structure construction and file I/O). End-to-end time is the entire program execution time for the analysis. For each case, we collect the results of our implementation, scaling up from 1--8 GPUs, and compare to the sequential Graph Challenge baseline. The least of 5 test runs is reported in the results. We present the analysis and end-to-end results in Figures \ref{fig:batch_analysis}, \ref{fig:batch_analysis_speedup} and \ref{fig:batch_total}, respectively, including batching variation in batch counts of 1, 5 and 10 (see \S\ref{ssec:method-batching}). We summarize our results as follows, followed by a discussion on the impact of concurrent batching. 

\textbf{Analysis Time:} 
In Figures \ref{fig:batch_analysis} and \ref{fig:batch_analysis_speedup}, we present execution time scaling per GPU count and overall performance improvement relative to the sequential GraphBLAS-based implementation. We observe the lowest execution time using a batch count of 10 and 8 GPUs of 1.17 seconds, compared to the sequential baseline of 64.23 seconds. At its peak, we observe a 55$\times$ performance improvement upon the sequential in analysis timing, with a geometric mean performance improvement at 8 GPUs of 47$\times$ across all three batch counts. We attribute performance improvements to improved data distribution using higher device counts, leveraging parallel performance.

Between batch counts, we observe the highest improvement using a batch count of 10, which noticeably outperforms the other batch counts across all GPU counts. On average, the larger batch count yields up to 170\% improvement in performance upon a batch count of 1 and up to 20\% at a batch count of 5. On average, using batching in this regard yields a 140\%improvement over the default case of a batch count of 1. In the single GPU case, the limitations in flexibility of even data distribution on device are very apparent in performance impacts. While the usage of higher device counts alleviates workloads for better distribution, the combination of batching and higher device counts yields the best flexibility for performant data distribution with lower data sizes at a given processing point.

\textbf{End-to-End Time:}
We present the end-to-end performance results in Fig.~\ref{fig:batch_total}, depicting the execution times with varied batch and GPU counts. Aside from the analytics time, we note that the data loading task is relatively expensive, taking approximately 40 seconds. Consequently, host to device data movement costs are nontrivial (about 100 seconds on a single GPU considering a single batch). Still, we demonstrate improvements relative to the sequential baseline by $\sim$4$\times$, with the highest improvement using a single batch at 8 GPUs by about 9$\times$. By geometric mean, we observe 8$\times$ execution time performance improvement across all data points. We further compare implicitly with the 2024 Graph Challenge champion \cite{sans}, with approximately 2$\times$ improvement in end-to-end execution time relative to their multithreaded implementation.

As shown in Fig.~\ref{fig:batch_total}, for \#GPU $>$3, a single batch outperforms that of 5 and 10 batch counts, while a batch count of 10 shows the best performance for \#GPUs $\le$ 3. This can be attributed to the relatively expensive data movement and initial container building costs combined with uneven workload distributions owing to the input data.  Similarly, we see minimal improvement using 10 batches over 5, with a 1.2$\times$ improvement and an average improvement with a batch count $>$1 of 1.1$\times$.

\textbf{Batching:} Relative to performance, batching serves as a means to balance workload distribution as GPU counts increase. In practice with C++26 \texttt{std::execution} device workload scheduling, resources appear to be scheduled efficiently relative to data sizes. Using the \texttt{nvexec} scheduler, if a workload fits in the capacity of a single GPU, resources are scheduled accordingly. This does not necessarily balance towards workloads, which is remedied through batching, providing a pre-scheduler workload distribution before bulk pushes to devices. This, alongside the sequential processing loop coupled with batching, results in variable performance improvements relative to tradeoffs in device counts and workload distributions. For our purposes, we express the usefulness of batching in large data instances with low device counts within the non-complex analytics workloads presented.

\begin{table}[ht]
\centering
\caption{Best packet rate (\emph{Higher} is \emph{Better}) per batch and \#GPUs.}
\label{tab:best-packet-rate}
\begin{tabular}{|c|c|c|}
\hline
Batch Count & Best Packet Rate (packets/s) & \#GPUs \\ \hline \hline
Sequential & 2,614,183 & - \\ \hline \hline
1 & \textbf{24,061,441} & 8 \\ \hline
5 & 23,626,502 & 8 \\ \hline
10 & 23,372,598 & 8 \\ \hline
\end{tabular}

\end{table}

\subsection{Packet Processing}
We further summarize our end-to-end execution time performance relative to the sequential in terms of network packets processed per second. This metric is calculated by recording the best end-to-end execution time relative to the number of packets in the randomized network traffic dataset. We highlight the best rate per batch count in Table \ref{tab:best-packet-rate}, observing approx. 9$\times$ improvement as compared to the reference baseline, with analysis times following at approximately 50$\times$ over the same.

\subsection{Observations}\label{ssec:productivity}

We discuss the observed productivity and challenges of C++26 \texttt{std::execution} model for multi-device execution.

\textbf{Productivity:} 
Primary benefit of the C++26 execution workflows is flexible device oriented programming through standardized solutions enhanced with vendor optimizations (e.g., device memory management via standardized containers and unified memory). Our implementation is approximately 20 lines of code (LoC) for each of the maximum scan and sum reduction operations, and in total about 100 LoC including the driver functions. We further note that our implementation includes \underline{no explicit CUDA code} and relies fully on vendor functionalities supplied within the standard. This is beneficial for free performance gains across GPU generations, without the need to adapt entire workflows to be device-aware, as long as the methods conform to a set of standard operations. Complex operations can be included through explicit device code within the lambdas.

\textbf{Challenges:}
Notable challenges exist in the conversion of complex algorithms to fit within the C++26 \texttt{std::execution} workflows, especially in the domain of graph problems. The straightforward bulk pushing of operations to devices limits user flexibility in controlling data distribution (which led us to consider batching), which is critical for irregular and hierarchical data instances. Furthermore, de facto unified memory requires careful examination of device data accesses for cache efficiency. These considerations, alongside those typical in parallel graph problems (e.g., load imbalance), still require interventions, as the scheduler does not act as an all-encompassing distributor across expansive problem domains.

\section{Concluding Remarks}
We utilize the recently proposed C++26 \texttt{std::execution} model on dense-GPU platforms, splitting the network traffic analytics workload into composable set of data manipulation operations as asynchronous task graphs. In addition, we show that explicit control of the data distributions across GPUs (i.e., batching) is still relevant for modern programming abstractions. Despite using high-level abstractions, we achieve up to 55$\times$ execution time improvement relative to the sequential linear-algebra baseline and about 2$\times$ improvement in implicit comparison of end-to-end execution time to the streaming-based multithreaded implementation by the 2024 Graph Challenge champion.


\section{Acknowledgments}
This work is in parts supported by the National Science Foundation under Grant No. 2047821, the DOE ASCR End-to-end co-design for performance, energy efficiency, and security in AI-enabled computational science (ENCODE) project at PNNL, and the Laboratory Directed Research and Development program at PNNL. PNNL is operated by Battelle Memorial Institute under Contract DE-AC05-76RL01830. We would also like to thank Dr. Tim Carlson at PNNL and the PNNL Research Computing staff for their continuous support.

\bibliographystyle{IEEEtranN}
\bibliography{refs}

\end{document}